# Recycling End–of–life Polycarbonate in Steelmaking; *Ab Initio* Study of Carbon Dissolution in Molten Iron

M. Hussein N. Assadi* and Veena Sahajwalla

*Centre for Sustainable Materials Research and Technology, School of Materials Science and Engineering, The University of New South Wales, Sydney, 2052, Australia*

Corresponding Author:
*Telephone: +61-2-9385-5234. E-mail: Hussein.assadi@unsw.edu.au

**ABSTRACT**: The scarcity of fossil fuels as carbon resources has motivated the steelmaking industry to search for new carbon sources such as end–of–life polymeric products. Using *ab initio* molecular dynamics simulation, we demonstrate that 41% of polycarbonate's carbon content is readily dissolved in molten iron's interface at $T = 1823$ K which is comparable to graphite with ~58% carbon content dissolution. More importantly, we demonstrate that polycarbonate's hydrogen content does not dissolve in molten iron but rather escape in gaseous form. Therefore, waste polycarbonate constitutes a feasible carbon source for steelmaking.

## 1. INTRODUCTION

Carbon is arguably the most important alloying element in steel that critically influences its mechanical,[1] magnetic[2] and corrosion properties.[3] Traditional carbon sources for carburizing steel have long been petroleum or metallurgical coke and various forms of natural graphite.[4] However, since the global crude steel production has been always increasing, reaching a new record of $1.518 \times 10^9$ tons/year in 2011,[5] a paramount pressure is straining the supply of traditional *carbonaceous materials* required to sustain the global steel production.[6] Therefore the steelmaking industry is thoroughly looking for innovative alternative carbon sources. In recent years end–of–life carbon–rich polymers[7,8] such as polycarbonate[9] have been experimentally investigated as a substitute or a supplement for traditional carbon sources. For instance, recent experiments showed 4.12% carbon pick–up within two minutes by molten iron from polycarbonate which exceeds the C pick–up from both coke and coal chars.[9] Furthermore, the use of the recycled polycarbonate as a carbon source is justified on environmental accounts as well. That is because polycarbonate is one of the fastest growing engineering plastics with broad applications in consumer goods. The annual global demand for polycarbonate exceeds ~3.41 million tons[10] while less than 20% of end–of–life polycarbonate products are currently being recycled.[11]

However, any further developments in utilizing polycarbonate as carbon source for steelmaking on industrial scale will require a critical feasibility assessment. For instance, the main concerns are the dissolution rate of carbon in molten iron from polycarbonate and the possible contamination of the final steel products by unintentional impurities such as H. To address these concerns, obtaining a detailed atomistic picture of the C dissolution process is indispensable. Nonetheless, due to arbitrary defined experimental settings and the macro-scale nature of the experiments, a full atomistic understanding of the dissolution mechanism is rather difficult to obtain experimentally.

On the other hand, the theoretical attempts by the means of classical molecular dynamics[12] and lattice based Monte–Carlo simulation,[13] so far, have only generated exiguous atomistic insight on the dissolution of carbon. These techniques have innate inadequacy in dealing with the diversity of the chemical species and the structural complexity involved in the dissolution process. These limitations are rooted in the lack of ternary and quaternary force–fields for molecular dynamics and the incapacity of lattice based Monte–Carlo simulation for liquid phase and interfacial interactions. Therefore, any successful theoretical techniques should be free of these limitations. In this regard *ab initio* molecular dynamic simulation based on density functional theory (DFT) emerges as a precious theoretical tool that circumvents the shortcomings of the previous methods. Recently *ab initio* molecular dynamics has been successfully applied to study both dissolution[14] and high temperature metallurgical[15] problems. Progressively, in this work we apply *ab initio* molecular dynamics simulation to investigate the dissolution of polycarbonate in molten Fe.

## 2. COMPUTATIONAL SETTINGS

Spin polarized *ab initio* molecular dynamics were performed with the Vienna *ab initio* simulation package VASP[16,17] utilizing plane-wave basis for expanding electronic wave functions[18] and projector augmented wave (PAW) method[19] to generate pseudopotentials. PAW pseudopotentials use smaller radial cutoffs for atomic core regions which results in more accurate and efficient calculations. The visualization and analysis of the outcomes was performed by VMD package.[20] Electronic exchange and correlation interactions were approximated by PW91 gradient corrected density functional.[21] The energy cutoff for all configurations was set 500 eV while only gamma point was used to generate the *k*-point grid. The self-consistency tolerance threshold was set at $10^{-5}$ eV/atom which adequately describes the dissolution process.[22] Convergence test was performed by choosing a denser of k-point grid of 2 × 2 × 1 and increasing the cut-off energy to 550 eV. Neither changes resulted in any noticeable change in the total energy. Molecular dynamics (MD) simulations were performed using Nose–Hoover thermostat[23,24] to produce an NVT ensemble. In all configurations the temperature was set to 1823 K, slightly higher than Fe's melting temperature. The time step for all molecular dynamics was set, at most, to be 0.5 fs. Choosing larger time steps would result in unphysical dynamics. This was because H possesses significantly smaller mass than the other elements and it would travel through repulsive potentials if time steps were set larger than 0.5 fs, rendering the convergence of self-consistent energy calculations impossible. MD runs were carried out up to 6 ps when the magnitude of the energy fluctuation became less than ∼0.1% of the total energy as presented in Figure 1(a). The dissolution rate of polycarbonate in the molten iron was analyzed by probing the element specific density profiles ($\rho(z)$) sampled over series of bins constructed in the supercell. These bins are set to be parallel to the molten iron surface e.g. $z$ direction. $\rho(z)$ is calculated using the following equation:

$$\rho(z) = \frac{\langle Nz \rangle}{A_{xy} \Delta_z} \quad (1).$$

Here, $\langle Nz \rangle$ is the time–averaged number of atoms in a given bin of the length of $\Delta_z$. In order to obtain meaningful spatial density profile, the bin size $\Delta_z$ should be chosen in a manner that allows the density variation within the bin to be far smaller than the density variation across neighboring bins. Therefore, we divided the supercell along $z$ direction into 250 bins for which $\Delta_z$ was equal to ∼0.1 Å. The time average was taken over 100 snapshots separated by 0.05 fs. The time interval between these snapshots was set an order of magnitude smaller than the MD simulation time steps to avoid averaging over the dissolution process.

### 3. SYSTEM SETTINGS

First, a supercell of $4a \times 4b \times 3c$ of conventional αFe cell containing 106 Fe atoms under three dimensional boundary conditions was constructed. Then this structure was allowed to equilibrate at 1823 K for 6 ps to reach the liquid phase. This configuration's pair correlation function $g(r)$ is presented in Figure 1(b) which is in good agreement with the experimental $g(r)$[25] and with the one obtained from classical molecular dynamics.[26] For instance, integrating the radial distribution function, that is $4\pi \rho r^2 g(r)$ ($\varrho$ is the atomic density calculated using the supercell parameters) to the first minimum yields a coordination number ($N_c$) of ∼13.0. This value for $N_c$ corresponds to the highest possible density of randomly packed atoms which is the case of liquid metals such as Fe.[16]

To study the dissolution of polycarbonate in molten iron, a new supercell based on the structure of molten Fe was constructed: first the output structure of the molten Fe was cleaved along $z$ direction then a vacuum slab of 15 Å was added to the cleaved surface to construct a supercell of the size of 11.47 Å × 11.47 Å × 23.54 Å with a molten Fe's surface being perpendicular to $z$ direction. Then one hydrogen terminated monomer of polycarbonate was placed at the interface of the molten Fe. This single monomer contained 16 C atom, 16 H atom and 3 O atoms. We examined few different initial configurations to investigate the effect of the initial coordination on the dissolution process. We found that an approximate distance of ∼1.5 Å between the surface of molten Fe and a parallel carbon loop in polycarbonate monomer, as demonstrated in Figure 2(a) best facilitates the dissolution process.

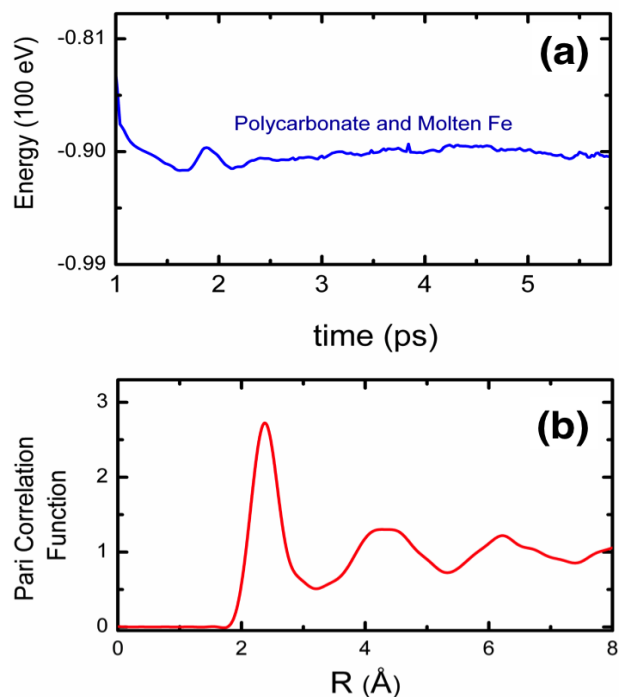

**Figure 1.** (a) Energy fluctuations during molecular dynamics simulation of interacting polycarbonate monomer and molten Fe system. (b) Pair correlation function of molten Fe at 1823 K obtained from *ab initio* molecular dynamics simulation.

## 4. RESULTS AND CONCLUSION

Figure 2(b) represents the equilibrium structure after the polycarbonate monomer interacted with the surface of the molten Fe. The charge density isosurface in Figure 2(b) indicates that at $T = 6$ chemical bonds break at multiple points throughout the decomposing monomer. Furthermore, Figure 2(b) also infers that only C atoms tend to dissolve into the molten Fe while no such tendency is observed for H and O atoms. A more detailed account for the interaction of different elements at the molten Fe's interface is provided by the density profiles in Figure 2(c) and (d). By comparing $\rho(z)$ at $t = 0$ and $t = 6$ ps, we can clearly identify the chemical decomposition process when the polycarbonate monomer loses its structural characteristic and collapses to smaller volume. Additionally, we can see that among all elements in the monomer, only C alone diffuses into the molten Fe. This is clearly deduced from Figure 2(d) where the H and O peaks are repulsed from the molten Fe surface while, contrarily, C peaks advance into the molten Fe. Consequently, 7 C atom of the original 16 C atoms were dissolved in the molten Fe by $t = 6$ ps. After initial C diffusion into the molten iron at the interface, dissolution process is facilitated by mass transfer phenomenon that leads to uniform carbon concentration at bulk level.[27] Additionally, the detached H atoms manifested in two forms: (i) gaseous hydrogen that is displaced away from the monomer appearing in the form of gaseous H peaks at $z = \sim 23$ Å and (ii) and individual H atoms that remained in the interface region. The produced gaseous hydrogen can be further utilized in steelmaking for indirect reduction of iron oxides and as a heat source.[28] Furthermore, Figure 2(d) also demonstrates that O atoms peaks overlapped with C peaks above the molten Fe interface indicating possible formation of CO or $CO_2$ gases.

We now define the dissolved C ratio in molten Fe as the number of C atoms that cross the interfacial line and dissolve in molten iron during the 6 ps time interval of the MD simulations. With $N_c$ representing the total number of C atoms in the supercell and $N_t$ the ones that have dissociated from polycarbonate and transferred into molten by the time $t = 6$ ps, the dissolved C ratio is $N_t/N_c$. By this definition we found that $\sim 41\%$ of the C content of the polycarbonate monomer is dissolved in molten iron which is comparable with the one of graphite that has been found to be $\sim 58\%$ (See Supporting Information). Here, we see that although the dissolved C ratio of the polycarbonate is 17% smaller than the one of graphite, recycled polycarbonate can be an economically plausible carburizing source as it is readily available as waste material.

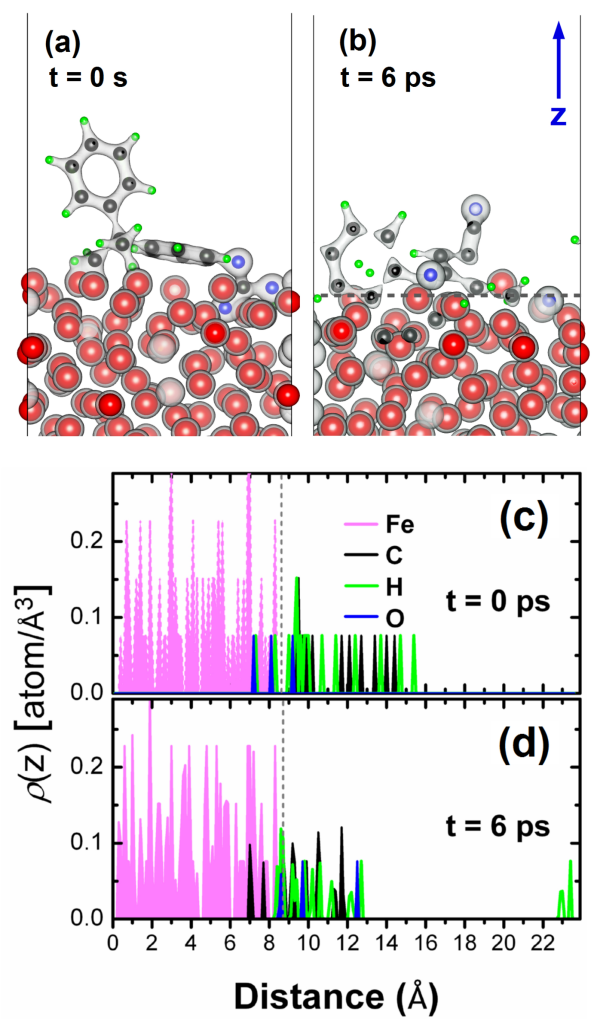

**Figure 2.** (a) Initial configuration of polycarbonate and molten iron surface. (b) Decomposition and dissolution of the polycarbonate monomer in molten iron at $T = 1823$ K after 6 ps. The presentation of atoms are as following: (i) Red spheres for Fe, (ii) black spheres for C, (iii) green spheres for H and (iv) blue spheres for oxygen. The gray shade surrounding the atoms represents the electronic charge density around each atom which also indicates the atomic bonds. (c) and (d) Calculated density profiles $\rho(z)$ of dissolution process of the polycarbonate monomer in molten Fe. The dashed line marks the interface defined as the highest $z$ coordinate of the Fe atoms. The isolated H peaks in (d) at $z = \sim 23$ Å denote molecular H that are detached from the monomer in gaseous phase.

Since the DFT–MD simulation can provide us with C dissolution ratio, we now attempt to calculate the *dissolution coefficient* of C from polycarbonate using the Noyes-Whitney equation:[29] $dC/dt = A(D/d)(\rho_s - \rho_b)$, in which $A$ is the surface area of the interface between carbonaceous material and the melt, $D$ is the diffusion coefficient, $d$ is thickness of the boundary layer at the interface, $\rho_s$ is the concentration of C on the surface and finally $\rho_b$ is the C concentration in the molten Fe's structure. We can calculate $D$ by equating $A$ to the cross–surface area of the supercell and $d$ to the volume slab with the height of 1.5 Å above the molten Fe where C–Fe interaction occur. Furthermore, $\rho_s$ and $\rho_b$ can readily be deduced from the C concentration in the interface and bulk areas. This procedure results in a C dissolution coefficient of 0.60 Å²/ps for polycarbonate. This is smaller than the C dissolution coefficient of graphite that was found to be 1.87 Å²/ps. When we consider that only ~81% of polycarbonate mass is C and its density is almost half of the density of graphite, one would have anticipated a smaller C diffusion rate for polycarbonate than the one of graphite. Nonetheless, when we consider that end–of–life polycarbonate is can be obtained at minimum cost, the calculated C diffusion rate seems satisfactory for carburizing molten Fe.

## CONCLUSION

By *ab initio* molecular dynamics simulation we reproduced the pair distribution function of molten Fe at 1823 K. We further demonstrated that polycarbonate has a C dissolution ratio of ~41% at molten iron's interface at this temperature. More importantly, in the case of polycarbonate, we found that, in the presence of C, H does not tend to dissolve in molten Fe. It rather either remains in the polycarbonate–melt interface or eventually escapes in gaseous form. Therefore, by offering exclusive C dissolution at a comparable rate to graphite, recycled end–of–life polycarbonate constitutes a feasible carbon source for steelmaking industry.

## ACKNOWLEDGEMENT

This work was supported by Australian Research Council's grant number FT0992021. The computational facility was provided by Intersect Australia Limited.

Supporting Information

# Recycling End–of–life Polycarbonate in Steelmaking; *Ab Initio* Study of Carbon Dissolution in Molten Iron

M. H. N Assadi* and Veena Sahajwalla

*Centre for Sustainable Materials Research and Technology, School of Materials Science and Engineering, The University of New South Wales, Sydney, 2052, Australia*

Corresponding Authors:
*Tel: +61-2-9385-5234. E-mail: Hussein.assadi@unsw.edu.au

To provide a comparative measure of polycarbonate's performance as a carbon source, we also studied the dissolution of graphite in molten Fe. This is because graphite is a well–studied[1] carburizing material and is known to have the highest carbon dissolution rate in molten Fe,[2] thus it sets a benchmark for polycarbonate's carburizing potentials. Two sheets of solid graphite with experimental lattice parameters[3] containing 48 C atoms were added at the distance of ~ 1.5 Å from the closest Fe atoms of the molten Fe's cleaved surface thus creating graphite–molten Fe interface as demonstrated in Figure S1 (a). Figure S1(b) shows the atomic configuration of dissolved graphite in molten Fe. We can see that the first layer of graphite has completely been diffused in the molten iron while the second layer has lost its crystalline nature. Figures S2(a) and (b) present the density profile of the graphite/molten Fe at the beginning and the conclusion of the simulation. We can see the two sharp C peaks in Figure 4(a) just above the graphite/molten Fe interface demonstrating the perfectly crystalline nature of the graphite. By $t$ = 6 ps, most of the C atoms that is 27 of the original 48 C atoms have been dissolved in the molted Fe by crossing the interface boundary.

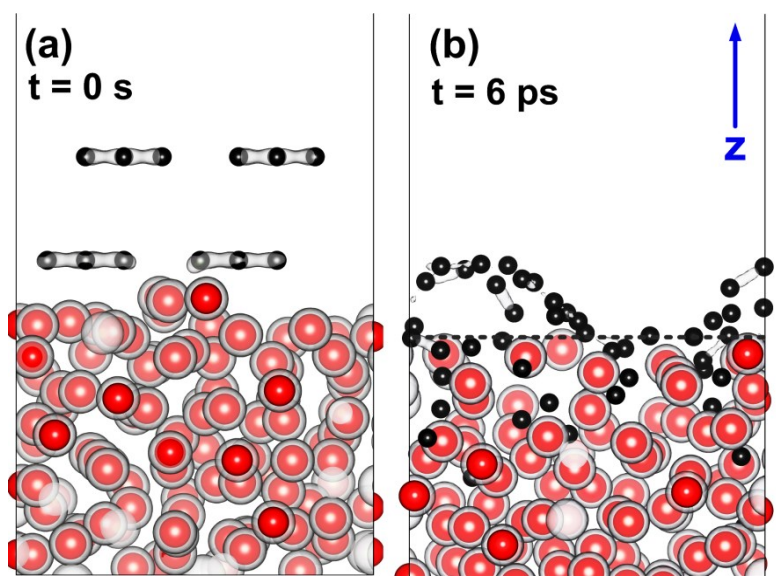

Figure S1. The decomposition and dissolution of the graphite sheets in molten iron at $T$ = 1823 K (~ 1550 °C). The red and black balls represent Fe and C atoms respectively. The gray shade surrounding the atoms represents electronic charge density around each atom which also indicates the atomic bonds. In (b) one can see the disappearance of C–C bonds as the graphite sheets are dissolved in molten Fe.

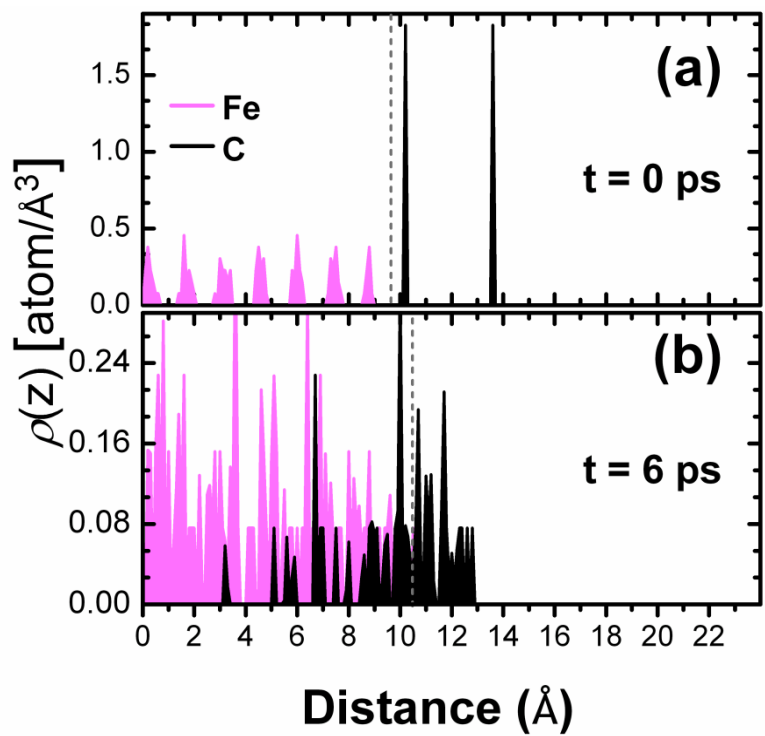

Figure S2. Calculated density profiles $\rho(z)$ of dissolution process of the graphite layers in molten Fe are presented. (a) and (b) correspond to the structures at the beginning and the conclusion of the simulation. The dashed line marks the interface defined as the highest $z$ coordinate of the Fe atoms.